\documentstyle[aps,multicol,epsf]{revtex}
\begin{document}
\draft
\title{Logarithmic Clustering in Submonolayer Epitaxial Growth}

\author{P. L. Krapivsky$^{1}$, J. F. F. Mendes$^{2}$, and S. Redner$^{1}$}

\address{$^{1}$Center for Polymer Studies and Department of Physics, 
Boston University, Boston, MA 02215 \\
$^{2}$Centro de F\'\i sica do Porto and Departamento de F\'\i sica, 
Universidade do Porto, 4150 Porto, Portugal}

\maketitle

\begin{abstract}

We investigate submonolayer epitaxial growth with a fixed monomer flux
and irreversible aggregation of adatom islands due to their effective
diffusion.  When the diffusivity $D_k$ of an island of mass $k$ is
proportional to $k^{-\mu}$, a Smoluchowski rate equation approach
predicts steady behavior for $0\leq\mu<1$, with the concentration $c_k$
of islands of mass $k$ varying as $k^{-(3-\mu)/2}$.  For $\mu\geq 1$,
continuous evolution occurs in which $c_k(t)\sim (\ln
t)^{-(2k-1)\mu/2}$, while the total island density increases as $N(t)
\sim (\ln t)^{\mu/2}$.  Monte Carlo simulations support these
predictions.

\end{abstract}


\begin{multicols}{2}

\narrowtext


Epitaxial thin film growth involves deposition of atoms onto a substrate
and diffusion of these adatoms, leading to their aggregation into
islands of ever-increasing size\cite{metois,ven}.  The resulting island
morphology and mass distribution depends intimately on the diffusion
processes of the adatoms.  While this connection has long been
recognized\cite{metois}, a complete understanding of this evolution is
still lacking.  For a variety of atomic transport mechanisms, there is a
power-law dependence of the effective island diffusivity $D_k$ on its
mass $k$, $D_k\propto k^{-\mu}$
\cite{morg,wen,pai,voter,sholl,khare,sol}, with $\mu$ typically in the
range $(1/2,3/2)$.  Starting with this observation, various approaches
have suggested that the growth of islands due to their clustering is a
power law in time\cite{pai,sholl}.

In this Letter, we provide a comprehensive account for the evolution of
the island size distribution in the submonolayer regime by solving the
Smoluchowski rate equations.  For mobility exponent $0\leq\mu<1$, a
steady state arises in which the concentration of islands of mass $k$ is
given by $c_k\propto k^{-\tau}$, with $\tau=(3-\mu)/2$.  For all
$\mu>1$, logarithmic island evolution occurs in which their total
density grows as $(\ln t)^{\mu/2}$ while $c_k(t)$ varies as $(\ln
t)^{-(2k-1)\mu/2}$.  More generally, our approach applies to {\em any}
epitaxial system in which the diffusivity of an island vanishes more
rapidly than inversely with its mass.

In the diffusion-controlled limit, the aggregation rate $K_{ij}$ of an
$i$-mer and a $j$-mer is given by the Smoluchowski formula $K_{ij}\sim
(D_i+D_j)(R_i+R_j)^{d-2}$\cite{ernst}.  Here $R_i$ is the linear size of
an $i$-mer (island of mass $i$), which is assumed to be compact, and $d$
is the spatial dimensionality of the substrate.  This Smoluchowski
formula is applicable in $d>2$, while in the physically relevant case of
two dimensions the reaction rate depends logarithmically on the island
size \cite{ernst}.  For both simplicity and because little quantitative
information is lost, we shall ignore these logarithmic factors.  This is
equivalent to treating the islands as point-like throughout their
evolution. 

With $D_k\propto k^{-\mu}$, an appropriate choice of time units, and the
neglect of logarithmic corrections, the reaction rate in two dimensions
becomes
\begin{equation}
\label{rate}
K_{ij}=i^{-\mu}+j^{-\mu},
\end{equation}
and the corresponding Smoluchowski rate equations for the evolution of
the concentrations of $k$-mers in the presence of a steady monomer flux
are
\begin{equation}
\label{smol}
{dc_k\over dt}= {1\over 2}\sum_{i+j=k}K_{ij}c_ic_j
-c_k\sum_{k=1}^\infty K_{kj}c_j+F\,\delta_{k1}.
\end{equation}
These rate equations represent a mean-field approximation in which
spatial fluctuations are neglected, and also a low-coverage
approximation, since only binary interactions are treated.  

Let us first consider the behavior in the steady state regime.  To solve
the rate equations in this case, we introduce two generating functions
\begin{equation}
\label{moments}
{\cal C}(z)=\sum_{k=1}^\infty c_k z^k, \quad
{\cal C}_\mu(z)=\sum_{k=1}^\infty k^{-\mu}c_k z^k.
\end{equation}
Multiplying Eq.~(\ref{smol}) by $z^k$, and summing over all $k$, gives
\begin{equation}
\label{mom}
{\cal C}_\mu(z){\cal C}(z)-{\cal C}_\mu(z)N
-{\cal C}(z)N_\mu + Fz=0.
\end{equation}
Here $N={\cal C}(z=1)=\sum c_k$ is the total island density
and $N_\mu={\cal C}_\mu(z=1)=\sum k^{-\mu}c_k$.

We now assume a power law asymptotic behavior for the steady state
concentration,
\begin{equation}
\label{asym}
c_k\simeq {C\over k^\tau}, 
\end{equation}
as $k\to\infty$.  For this power law to hold for all $k$, we require
$\tau>1$, so that $\sum k^{-\tau}$ converges; this leads to the
condition $\mu<1$ for the mobility exponent as shown below.  From basic
Tauberian theorems\cite{hardy}, the asymptotic form for $c_k$ in
Eq.~(\ref{asym}) induces the following power-law singularities in the
generating functions as $z\to 1$
\begin{eqnarray}
\label{gen}
{\cal C}(z) &=&N+C\Gamma(1-\tau)(1-z)^{\tau-1}+\ldots,\nonumber \\
{\cal C}_\mu(z)&=&
N_\mu+C\Gamma(1-\tau-\mu)(1-z)^{\tau+\mu-1}+\ldots.
\end{eqnarray}
The leading constant factor in each line is finite and coincides with
the definition given in Eq.~(\ref{moments}) if the exponent of the
second term is positive.  Otherwise, the constant factor vanishes and
the generating function has a power-law divergence as $z\to 1$.
Substituting these expansions into Eq.~(\ref{mom}) and matching the
leading behavior in $(1-z)$ leads to the decay exponent
$\tau=(3-\mu)/2$.  The condition for a steady state to occur, $\tau>1$,
thus imposes an upper bound on the mobility exponent, $\mu<1$.  From
matching the leading behavior in $(1-z)$, the constant $C$ may also be
determined, from which the island mass distribution in the steady-state
regime $0\leq\mu<1$ is
\begin{equation}
\label{steady}
c_k\simeq \sqrt{{F\over 4\pi}(1-\mu^2)\cos(\pi\mu/2)}
\,\,k^{-(3-\mu)/2}.
\end{equation}

It is important to note that this mass distribution holds only up to a
mass cutoff $k_c(t)\sim t^\zeta$ whose value is determined by requiring
that the total mass in the system due to the steady input is
proportional to $t$ -- islands of mass greater than $k_c(t)$ have not
yet formed.  Therefore
\begin{equation}
\label{zeta}
M(t)=\sum_{k=1}^\infty kc_k(t)\sim 
\sum_{k=1}^{k_c}k^{(\mu-1)/2}\sim t^{(\mu+1)\zeta/2}\sim t,
\end{equation}
which gives the mass cutoff exponent 
\begin{equation}
\label{zet-exp}
\zeta(\mu)=2/(\mu+1).
\end{equation}

We now investigate the asymptotic behavior of the island mass
distribution for $\mu\geq 1$.  In the extreme case of $\mu=\infty$, {\it
i.e.}, diffusing monomers and {\em immobile} islands, the island density
grows as a power law in time, $N(t)\simeq (3F^2t)^{1/3}$
\cite{paul,bla,bartelt}.  We shall argue that continuously evolving
behavior occurs for all $\mu\geq 1$, but with anomalously slow
logarithmic kinetics.  When $\mu$ is strictly greater than unity but
still finite, we find
\begin{equation}
\label{log}
N(t)\simeq \sqrt{F}\left[{\sin(\pi/\mu)\over\pi}\,\,\ln T\right]^{\mu/2}, 
\end{equation}
with $T\equiv t\sqrt{F}$, while the concentration of $k$-mers decays in
time as
\begin{equation}
\label{clog}
c_k(t)\sim \sqrt{F}\,(k!)^\mu\,
\left(\ln T\right)^{-\mu(2k-1)/2}.
\end{equation}
It is remarkable that such logarithmic dependences, a feature which
generally signals marginal behavior, occurs in the entire regime
$1<\mu<\infty$.  In the borderline case of $\mu=1$, even more unusual
behavior arises with $N(t)\sim \sqrt{\ln T/\ln(\ln T)}$.

Our argument leading to Eqs.~(\ref{log}) and (\ref{clog}) is based on a
quasi-static approximation, in which the time derivative in
Eq.~(\ref{smol}) is neglected.  Indeed, the logarithmic behavior in
Eqs.~(\ref{log}) and (\ref{clog}) immediately implies that the temporal
derivatives in the Smoluchowski rate equations are asymptotically
negligible.  Within this quasi-static framework, Eqs.~(\ref{smol}) become
\begin{eqnarray}
\label{sm}
0&=&1-c_1\left(N+N_\mu\right), \nonumber \\
0&=&\frac{1}{2}\sum_{i+j=k}\left({i^{-\mu}}+{j^{-\mu}}\right)c_ic_j
   -c_k\left(k^{-\mu}N+N_\mu\right).
\end{eqnarray}
Further, by summing Eqs.~(\ref{sm}) over all $k$, the total island
density in the quasi-static limit obeys
\begin{equation}
\label{nqs}
0=1-NN_\mu
\end{equation}
In Eqs.~(\ref{sm}) and (\ref{nqs}) we have set $F=1$ by a rescaling of
units.  Eq.~(\ref{nqs}) immediately gives $N_\mu=N^{-1}$, and then from
the first of Eqs.~(\ref{sm}), $c_1\simeq 1/N$.  The remainder of
Eqs.~(\ref{sm}) may then be solved recursively.  By writing the first
few of these equations, it is evident that the dominant contribution to
$c_k$ is the term in the quadratic product which is proportional to
$c_1c_{k-1}$.  If we keep only this contribution, the resulting
recursion may be solved straightforwardly to yield
\begin{eqnarray}
\label{soln}
c_k&\simeq&{1\over N}\,\prod_{j=2}^k
(1+N^2j^{-\mu})^{-1}\,
\prod_{j=1}^{k-1}\left(1+j^{-\mu}\right) \nonumber\\
&\equiv&{1\over N}\prod_{j=2}^k B_j\prod_{j=1}^{k-1} b_j.
\end{eqnarray}
Since the factors $B_j\ll 1$ for $j^\mu\ll N^2$, while $B_j\to 1$ for
$j^\mu\gg N^2$, this implies that $c_k$ is a rapidly decreasing function
of $k$ for $k\ll N^\mu$ and then becomes constant for larger $k$.

To compute $c_k$, first note that for $\mu>1$ the product $\prod_j b_j$
converges, so that it may treated as constant.  We then write the second
product as the exponential of a sum and take the continuum limit.  This
leads to
\begin{eqnarray}
\label{ck-cont}
c_k&\sim&{1\over N}\exp\left
[-\sum_{j=2}^k\ln (1+N^2j^{-\mu})\right]\nonumber\\
 &\simeq& {1\over N}\exp\left[-N^{2/\mu}\int_0^x\ln(1+w^{-\mu})\,dw\right],
\end{eqnarray}
where $w=j/N^{2/\mu}$ and $x=k/N^{2/\mu}$.  This form has two slightly
different asymptotic behaviors depending on whether $\mu$ is strictly
greater than or equals 1.  For $\mu>1$, the monotonically increasing
integral in Eq.~(\ref{ck-cont}) converges as $x\to\infty$.  Thus $c_k$
decreases as a function of $k$ until a threshold value $k_{\rm th}\simeq
N^{2/\mu}$, beyond which $c_k$ remains constant with a value determined
by taking the upper limit of the integral as infinite.  Hence
\begin{equation}
\label{cstar}
c_{\rm th}\sim {1\over N}\exp\left[-A_\mu\,N^{2/\mu}\right], 
\end{equation}
with $A_\mu=\int_0^\infty\ln(1+w^{-\mu})\,dw=\pi/\sin(\pi/\mu)$.
Physically, we expect this constancy in $c_k$ to persist until $k$
reaches the cutoff $k_c\sim t^\zeta$.

To check this result, we performed numerical simulations in the mean
field limit of submonolayer epitaxial growth.  In the simulation, an
island of mass $k$, which remains point-like throughout the aggregation
process, moves equiprobably to any site with a probability proportional
to $k^{-\mu}$, as mandated by the power-law mass-dependent island
diffusivity.  There is also a steady monomer flux entering the system.
As shown in Fig.~1, $c_k(t)$ is nearly constant in $k$ over a
substantial range as predicted by our theory.  However, when $k\approx
k_c$, there is a peak in $c_k$ which is not accounted for in our
quasi-static description.
   
\begin{figure}
\epsfxsize=70mm
\epsfysize=70mm
\centerline{\epsffile{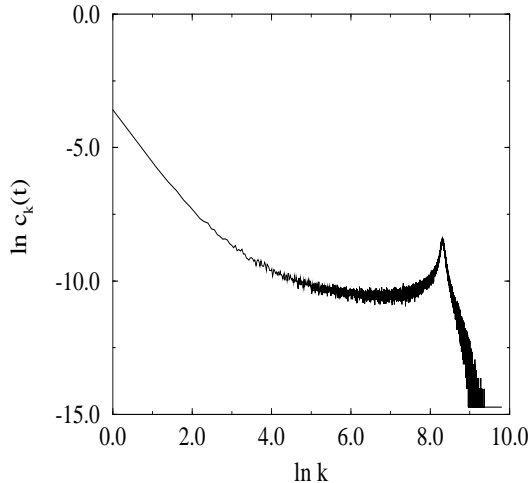}}
\caption{$c_k(t)$ versus $k$ on a double logarithmic scale at
$t\approx 22000$ for $\mu=1.5$.  The data is based on 5000 realizations
of an initially empty system with $F=0.05$.}
\end{figure}

\begin{figure}
\epsfxsize=70mm
\epsfysize=70mm
\centerline{\epsffile{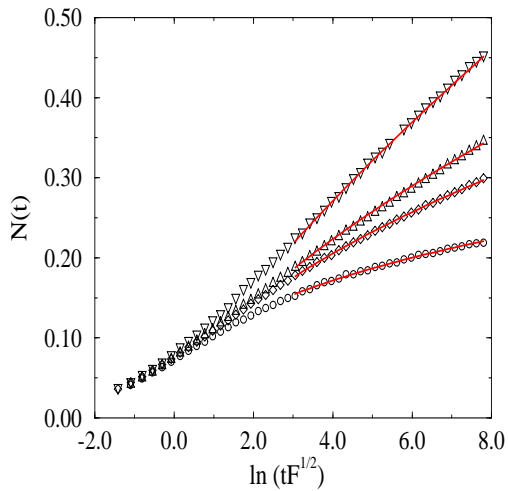}}
\caption{$N(t)$ versus $\ln(tf^{1/2})$ for $\mu=1.2$ ($\circ$), 1.4
($\diamond$), 1.5 ($\bigtriangleup$), and 2.0 ($\bigtriangledown$).  The
data is based on 1000 realizations.  Also shown are power law fits to
the data in the range $t>3$.  This gives, for the exponent of $\ln t$,
0.38, 0.55, 0.64 and 0.76, respectively, for $\mu=1.2$, 1.4, 1.5, and
2.0.}
\end{figure}

The time dependence of the total island concentration may now be
determined by using the fact that the mass density in the system grows
linearly with time, $M(t)=\sum_{k\geq 1} kc_k(t)=t$.  The dominant
contribution to this sum is from the plateau region $k_{\rm th}<k<k_c$,
where $c_k$ is approximately constant in $k$, $c_k\approx c_{\rm th}$.
Thus
\begin{equation}
\label{loga}
M(t)\sim c_{\rm th}\sum_{k=k_{\rm th}}^{k_c}k
\sim c_{\rm th} k_c^2\sim t.
\end{equation}
Here we also use the fact that $k_{\rm th}$ grows only logarithmically
in time (see below), so that the lower limit can be taken to be zero.
Furthermore, $k_c(t)\sim t^\zeta$ grows more rapidly than $t^{1/2}$.
Indeed, since the cutoff exponent $\zeta(\mu)=2/(\mu+1)$ for $\mu<1$
(Eq.~(\ref{zet-exp})) and $\zeta(\mu=\infty)=2/3$
\cite{paul,bla,bartelt}, we anticipate that $2/3\leq \zeta(\mu)\leq 1$
when $1\leq\mu<\infty$.  In fact, $\zeta=1$ in this range of mobility
exponent, as we will show below.  Using $\zeta=1$, Eq.~(\ref{loga}),
together with $c_{\rm th}\sim \exp(-A_\mu N^{2/\mu})$ and $k_c\sim
t^\zeta$, we immediately obtain Eq.~(\ref{log}).  Our data for $N(t)$ as
a function of time (Fig.~2) is qualitatively consistent with $N(t)$
growing as a power of $\ln t$, but with a smaller exponent than $\mu/2$.
Note finally that $k_{\rm th}\sim N^{2/\mu}$ which is proportional to $\ln t$.

To determine $c_k(t)$ for $k^\mu\ll N^2$, we now use the approximation
$B_j\approx j^{\mu}/N^2$ in Eq.~(\ref{soln}) to give
\begin{equation}
\label{ckvsn}
c_k\sim {(k!)^\mu\over{N^{2k-1}}},
\end{equation} 
which directly leads to Eq.~(\ref{clog}).  Finally, the time dependence
of $c_{\rm th}(t)$ and $k_c(t)$ may be determined from the sum rules
(\ref{loga}) and $\sum c_k \sim c_{\rm th} k_c\sim N$.  These
two relations give $c_{\rm th}\sim N^2/t$ and $k_c\sim t/N$.  Thus in
the plateau region $k_{\rm th}<k<k_c$,
\begin{equation}
\label{ck-p}
c_k(t)\sim  {(\ln t)^\mu\over t},\qquad
k_c(t)\sim {t\over{(\ln t)^{\mu/2}}}.
\end{equation}

It is important to note that our approach applies to any mass-dependent
island diffusivity which decays faster than its inverse mass.  For this
general situation, the analog of Eq.~(\ref{soln}) is $c_k \sim N^{-1}
\prod^k (1+D_j) (1+N^2D_j)^{-1}$.  For example, for $D_k\sim e^{-2ak}$,
a case that was considered numerically in\cite{kuip}, we obtain
$N(t)\sim \exp(\sqrt{a\ln t})$.  This unusual growth -- faster than any
power of logarithm but slower than any power law -- would be difficulty
to observe numerically.

In the specific case $\mu=1$, subtler nested logarithmic behavior
arises, as reflected by the additional singularity in Eq.~(\ref{log}) as
$\mu\to 1$.  First, the product $\prod_{j=1}^{k-1} b_j=\prod_{j=1}^{k-1}
(1+j^{-1})$ in Eq.~(\ref{soln}) now equals $k$.  Second, the term
$c_2c_{k-2}$ also contributes to the asymptotic behavior. Third, and
most importantly, the integral in Eq.~(\ref{ck-cont}) diverges at the
upper limit.  Due to the second attribute, the recursion relation for
$c_k$ becomes
\begin{equation}
\label{ck-mar}
{c_k\over k}\,{1+k/N^2\over k/N^2 }={c_{k-1}\over k-1}
+{c_{k-2}\over k-2}\,{1\over N^2}.
\end{equation}
We seek a solution for $c_k$ in the form of Eq.~(\ref{soln}).  Thus we
write
\begin{equation}
\label{ck-marg}
c_k\sim {\cal C}_k\,{k\over N}\prod_{j=2}^k (1+N^2j^{-1})^{-1},
\end{equation}
where the factor ${\cal C}_k$ accounts for the additional term in
Eq.~(\ref{ck-mar}).  Substituting into Eq.~(\ref{ck-mar}) gives
\begin{equation}
\label{Bk}
{\cal C}_k={\cal C}_{k-1}
+{\cal C}_{k-2}\left({1\over k-1}+{1\over N^2}\right).
\end{equation}
These coefficients are slowly varying in $k$ when $k\gg 1$ and we may
treat $k$ as continuous in this asymptotic regime.  Eq.~(\ref{Bk}) then
becomes a differential equation whose solution is ${\cal C}_k\sim
k\,e^x$ (with $x=k/N^2$).  Consequently,
\begin{equation}
\label{ck-margin}
c_k\sim {k^2\over N}\,\exp\left[x
-N^2\int_0^x \ln\left(1+w^{-1}\right)\,dw\right].
\end{equation}
Thus for $\mu=1$, $c_k$ decreases rapidly in $k$ for $k\ll N^2$ and then
increases until $k$ reaches $k_c$.  The island mass distribution attains
a minimum at $x_{\rm th}=N^2$ whose value is
\begin{equation}
\label{ck-star-marg}
c_{\rm th}\sim \exp\left[-N^2\ln N^2\right].
\end{equation}

Paralleling the analysis of the case $\mu>1$, the total island
concentration is
\begin{equation}
\label{2log}
N(t)\sim \sqrt{\ln t\over \ln(\ln t)},
\end{equation}
while Eq.~(\ref{ck-marg})
together with ${\cal C}_k\sim k$ implies that the concentration of
islands for mass $k\ll N^2$ is
\begin{equation}
\label{3log}
c_k\sim {(k+1)!\over N^{2k-1}}
\sim (k+1)!\,\left[{\ln(\ln t)\over \ln t}\right]^{k-1/2}.
\end{equation}
Finally from Eq.~(\ref{ck-p}), the time dependence
of $c_k(t)$ and $k_c(t)$ is given by
\begin{equation}
\label{ck-p-mu}
c_k(t) \sim    {1\over t}\,{\ln t\over \ln(\ln t)},\qquad
k_c(t)    \sim    t\,\sqrt{{\ln(\ln t)\over \ln t}} 
\end{equation}

Our results should be generally applicable to real epitaxial systems in
the submonolayer regime.  This regime requires $Ft \ll 1$, while the
asymptotic predictions of our Smoluchowski theory apply for
$t\sqrt{F}\gg 1$.  Consequently, our results should be valid for
$F^{-1/2}\ll t\ll F^{-1}$.  Since the dimensionless flux $F$ is small in
typical epitaxy experiments, the time range over which our theory will
apply is correspondingly large.  Notice that the maximum island density
attained at the end of the submonolayer regime $t_{\rm max}\sim F^{-1}$
scales with flux as $N_{\rm max}\sim F^{1/2}[\ln(1/F)]^{\mu/2}$.  In
fact, for all systems with the diffusivity of an island decaying more
rapidly than its inverse mass, our approach leads to $N_{\rm max}$
universally being proportional to $F^{1/2}$ times a subdominant
model-dependent factor.

In conclusion, we determined the kinetics of islanding in submonolayer
epitaxial growth, in which adatom hopping induces a power-law
mass-dependent island diffusion, with $D_k\propto k^{-\mu}$.  This leads
to the reaction rate $K_{ij}\propto (i^{-\mu}+j^{-\mu})$ between two
islands of mass $i$ and $j$ in the Smoluchowski rate equation.  A steady
state arises for mobility exponent $0\leq \mu<1$, in which the island
concentration varies as $c_k\sim k^{-(3-\mu)/2}$ for $k<k_c\propto
t^\zeta$, with $\zeta(\mu)=2/(1+\mu)$.  Strikingly, logarithmic time
dependence arises for all $1<\mu<\infty$, a feature suggestive of
marginal behavior over this entire range.  In this regime, the total
island density $N(t)$ grows as $(\ln t)^{\mu/2}$, while the density of
islands of mass $k$ is $c_k(t)\propto (\ln t)^{-(2k-1)\mu/2}$, for $k\ll
k_{\rm th}\propto\ln t$, $c_k(t)$ independent of $k$, with $c_k(t)\sim
t^{-1}(\ln t)^\mu$, for $k_{\rm th}\leq k\leq k_c$, and $c_k$
vanishingly small for $k>k_c$.  For $\mu=1$, even more unusual nested
logarithmic behavior occurs.

JFFM gratefully acknowledges support from Funda\c c\~ao Luso Americana
para o Desenvolvimento (FLAD), and JNICT/PRAXIS XXI: grant /BPD/6084/95
and project PRAXIS/2/2.1/Fis/299/94.  PLK and SR gratefully acknowledge
support from NSF grant DMR9632059 and ARO grant DAAH04-96-1-0114.

\end{multicols}
\end{document}